\documentclass[letterpaper]{aa}
\usepackage{graphicx}

\def \ob{XMMU\,J174554.4$-$285456}

\begin{document}

\title{Discovery of a bright X-ray transient in the Galactic Center with XMM-Newton}

\author{D. Porquet\inst{1} 
           \and N. Grosso\inst{2}
           \and V. Burwitz\inst{1}
            \and I.L. Andronov\inst{3,4}
            \and B. Aschenbach\inst{1}
            \and P. Predehl\inst{1}  
            \and R.S. Warwick\inst{5}
       }
\offprints{Delphine Porquet\\ (dporquet@mpe.mpg.de)}

\institute{Max-Planck-Institut f\"{u}r extraterrestrische Physik,
P.O. Box 1312, Garching bei M\"{u}nchen D-85741, Germany
\and Laboratoire d'Astrophysique de Grenoble, Universit\'e Joseph-Fourier,
 BP53, 38041 Grenoble Cedex 9, France
\and Department of Astronomy, Odessa National University, T. G. Shevchenko park,
 65014 Odessa, Ukraine
\and Crimean Astrophysical Observatory, 98409 Nauchny, Ukraine
\and Department of Physics and Astronomy, University of Leicester,
 Leicester LE1 7RH, UK
}
\date{Received ... ; Accepted ...  }

\abstract{We report the discovery of a bright X-ray transient object, \object{\ob},  
observed in outburst with {\sl XMM-Newton} on October 3, 2002,
and located at 6.3$^{\prime}$ from \object{Sgr\,A*},
the supermassive black hole at the Galactic center. 
This object exhibits a very large X-ray luminosity variability
 of a factor of about 1\,300
  between two X-ray observations separated by four months. 
The X-ray spectrum is best fitted by a power-law with a photon index of 1.6$\pm$0.2
and absorption column density of 14.1$^{+1.6}_{-1.4}\times$10$^{22}$\,cm$^{-2}$. 
This large absorption suggests this source is located at the distance of the 
Galactic center, i.e., 8\,kpc. The 2--10\,keV luminosity is about
$1.0 \times 10^{35}\,(d/{\rm 8\,kpc})^{2}$\,erg\,s$^{-1}$. 
 A pulsation period of about 172\,s is hinted by the timing analysis.
The X-ray properties strongly suggest a binary system with either a black hole or 
a neutron star for the compact object.

}
\titlerunning{Discovery of a bright X-ray transient in the GC region}
\authorrunning{Porquet et al.}
\maketitle

\section{Introduction}

The Galactic center region ($\Delta l\sim2^\circ$, $\Delta b \sim 0.5^\circ$)  
is very complex in X-rays with both diffuse emission and point-like sources, 
counterparts of fluorescent molecular clouds, supernova remnants, compact objects,
 and stellar clusters. In addition to Sgr\,A*, the supermassive black hole at the
 Galactic center, this region shelters accreting compact objects, such as neutron stars
 and black hole candidates (e.g., Churazov et al.\ \cite{Ch97}, 
Sidoli et al.\ \cite{Si99}, Sakano et al.\ \cite{Sa02}, Porquet et al.\ \cite{P03a}),
 which can be transient sources in X-rays. 
Therefore, repeated X-ray observations of this region, achieved for the monitoring of 
Sgr\,A*, give us the opportunity to catch one of these compact objects during an 
outburst phase, and to shed light on their nature.  
We report here the serendipitous discovery on October 3, 2002 with {\sl XMM-Newton}
 of a bright X-ray transient source located at 6.3$^{\prime}$ from Sgr\,A*. 

\section{XMM-Newton observation}
\begin{figure}[!b]
\centerline{\includegraphics[width=7.8cm]{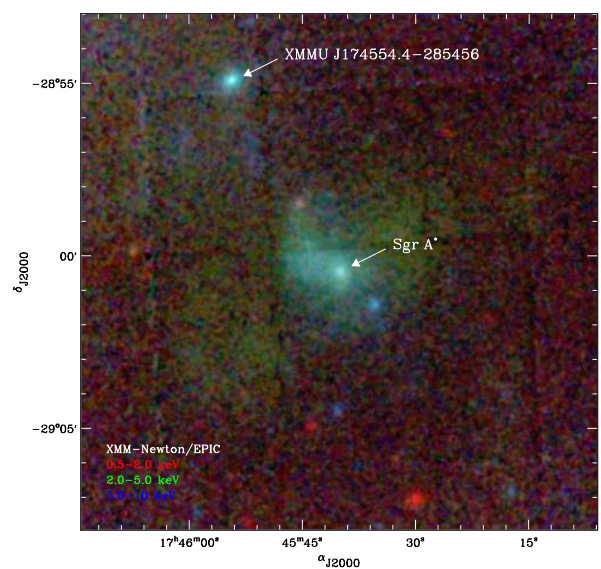}}
\vspace*{-0.2cm}
\caption{XMM-Newton/EPIC image of the bright X-ray transient \ob~ close
to Sgr\,A*. Red, green and blue code for photons with energy 0.5--2, 2--5, and 5--10\,keV,
 respectively.}
\label{fig:ima}
\vspace*{-0.3cm}
\end{figure}
The {\sl XMM-Newton} observation is the one obtained on October 3, 2002, 
where a very bright X-ray flare was reported from Sgr\,A*
 (Porquet et al.\ \cite{P03b}).
The observation exposure times are $\sim$16.8\,ks and $\sim$13.9\,ks for 
the MOS and  PN  cameras, respectively. The data were processed with 
{\tt SAS} (version 6.0). 
X-ray events with patterns 0--12 and 0--4 are used for MOS and PN, respectively. 
We select only events with data quality flag equal to 0. 
The astrometry is the one reported in Porquet et al.\ (\cite{P03b}). 
Fig.~\ref{fig:ima} shows the $7\arcmin\times7\arcmin$ central part of this observation 
in the 0.5--10\,keV energy range: 
a new X-ray source, as bright as Sgr\,A*, is clearly seen at 6.3$\arcmin$ from Sgr\,A*.
Its position in J2000 coordinates is $\alpha$=17h45m54.4s,
 $\delta$=$-28^\circ54\arcmin56\arcsec$
 with a 90$\%$ confidence level error position of 2.3$^{\prime\prime}$ in radius.
We name this new X-ray source \ob. We did not find any known counterpart of this
 source in the SIMBAD data base. We note that this area was previously observed
 in X-rays with higher sensitivity both with {\sl XMM-Newton}
 (e.g., Sakano et al. \cite{Sa04}) 
and {\sl Chandra} (e.g., Muno et al.\ \cite{Mu2003}), and that this source was not
 detected.  In addition, during the {\sl Chandra} observation on June 19, 2003,
 about eight months after the present observation, the source was not detected,
 hence \ob~is a transient X-ray source.
\begin{figure}[!t]
\centerline{\includegraphics[width=0.65\columnwidth,angle=90]{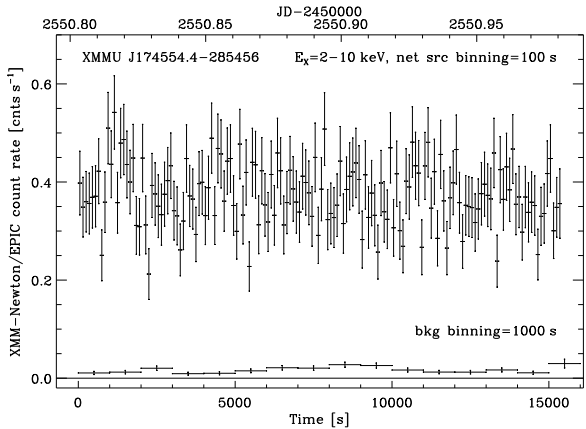}}
\vspace*{-0.1cm}
\caption{XMM-Newton/EPIC background subtracted light curve of \ob~
 in the 2--10\,keV energy band. The lower light curve shows the background level
 scaled to the source extraction area.}
\label{fig:lc}
\vspace*{-0.1cm}
\end{figure}

We extract the source events from a circular region of 12$^{\prime\prime}$-radius
 centered on the X-ray source position, and the background events from a rectangular
 region free of X-ray sources lying on the same CCD and close to the transient.
 Photon arrival times were computed for the solar system barycenter.
Fig.~\ref{fig:lc} displays the background subtracted EPIC light curve of \ob~ in the
 2--10\,keV energy band, during the time interval when the three cameras were
 observing together. We do not see any obvious variations of the light curve as
 type-I burst, or eclipse.

\section{Spectral analysis}\label{sec:spectral}

\begin{figure}[h!]
\vspace*{0.3cm}
\centerline{\includegraphics[width=0.6\columnwidth,angle=-90]{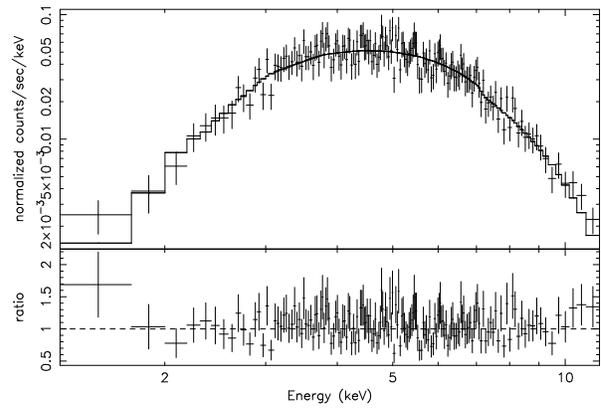}}
\caption{PN spectrum of \ob~with the best fit model using an absorbed power-law continuum (see Table~\ref{tab:fit}) and data-to-model ratio.}
\label{fig:PN}
\vspace*{-0.1cm}
\end{figure}

We use for the spectral analysis only PN data, which have a higher
S/N and better energy coverage than the MOS data.  
The spectrum is binned to a minimum of 20 counts per bin. 
The fitting parameter errors quoted correspond to 90$\%$
confidence ranges for one interesting parameter
($\Delta \chi^{2}$=2.71). 
In the following, we use the updated X-ray absorption cross-sections 
and abundances of the interstellar medium (ISM) of Wilms et al.\ (\cite{Wi00}). 
The PN spectrum, shown in Fig.~\ref{fig:PN}, is strongly absorbed below 2\,keV, 
evidence of a large absorption by the ISM along the line-of-sight.  
We fit the data taking into account the scattering of X-rays by dust, 
using the {\tt scatter} model (Predehl \& Schmitt \cite{PS95}) assuming a visual extinction value 
$A_{\rm V}$=30\,mag, as determined from IR observations of stars close to Sgr\,A*
 (e.g., Rieke et al.\ \cite{Rieke89}).

\begin{table}[!t]
\caption{Models of the PN spectrum of \ob, taking into account absorption.
 {\tt po}: power law.
{\tt bb}: black body. {\tt diskbb}: multi black body 
 (Mitsuda et al.\ \cite{M84}).  {\tt brems}: bremsstrahlung. 
 The number of degrees of freedom is 161. 
The column density is expressed in units of 10$^{22}$\,cm$^{-2}$.
The flux ($F_{\rm X}$) and luminosity ($L_{\rm X}$) corrected for absorption 
are expressed in 10$^{-11}$\,erg\,cm$^{-2}$\,s$^{-1}$ and 10$^{35}$\,erg\,s$^{-1}$, respectively. 
The luminosity is calculated assuming $d$=8\,kpc.}
\begin{tabular}{@{}cccccccc@{}}
\hline
\hline
\smallskip
Model  &   ${\cal N}_{\rm H}$ &  $\Gamma$ or $kT$ & $\chi^{2}_{\rm red}$ & $F_{\rm X}$ & $L_{\rm X}$ & $F_{\rm X}$ & $L_{\rm X}$ \\
       &  &  & & \multicolumn{2}{c}{2--10\,keV} & \multicolumn{2}{c}{0.5--10\,keV} \\
\hline
{\tt po}     &  14.1$^{+1.6}_{-1.4}$  & 1.65$^{+0.17}_{-0.16}$ & 0.87 & 1.4 & 1.0 & 2.1 & 1.5 \\
{\tt bb}     &  7.3$\pm$0.9  & 1.9$\pm$0.1      &1.08&   1.0 & 0.7  & 1.0 & 0.7    \\
{\tt diskbb} &  11.0$\pm$1.0  & 3.7$\pm$0.5     & 0.93 & 1.2 & 0.9 & 1.4 & 1.0 \\
{\tt brems} &   13.6$\pm$1.2  & 22$^{+40}_{-15}$  & 0.88 & 1.3 & 0.9 & 1.8 & 1.3 \\
\hline
\end{tabular}
\label{tab:fit}
\vspace*{-0.3cm}
\end{table}

The PN spectrum is well fitted by the standard absorbed continuum models, 
such as power-law, black-body, multi black body, and bremsstrahlung (Table~\ref{tab:fit}). 
However, the temperatures found for the latter three models are rather high, 
and the spectrum is not well fitted above 10\,keV by the black-body and the multi black body models. 
According to Predehl \&  Schmitt (\cite{PS95}), $A_{\rm V}$=30\,mag   
corresponds to an hydrogen column density of about $6\times10^{22}$\,cm$^{-2}$
towards the Galactic center located at about 8\,kpc. 
 The updated ISM abundance used here lead to absorption column density 
values about 30$\%$ larger than those derived assuming solar abundance
 as in Predehl \&  Schmitt (\cite{PS95}).   
For comparison we found for Sgr\,A* (during its X-ray flare) 
a column density of about 20$\times$10$^{22}$\,cm$^{-2}$ (Porquet et al. \cite{P03b}).
\begin{figure*}[!ht]
\begin{tabular}{cc}
\includegraphics[width=0.7\columnwidth,angle=90]{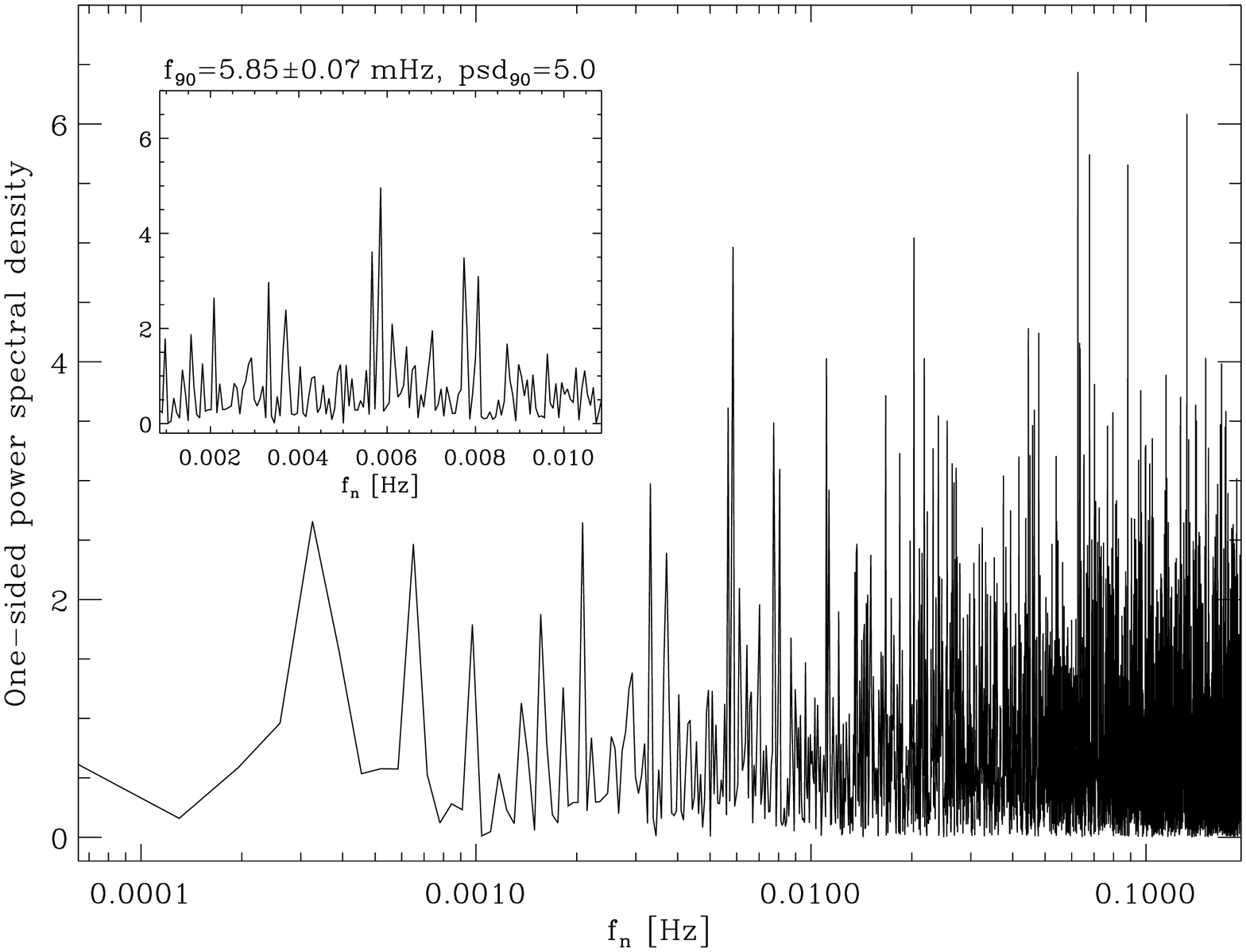} & \includegraphics[width=\columnwidth]{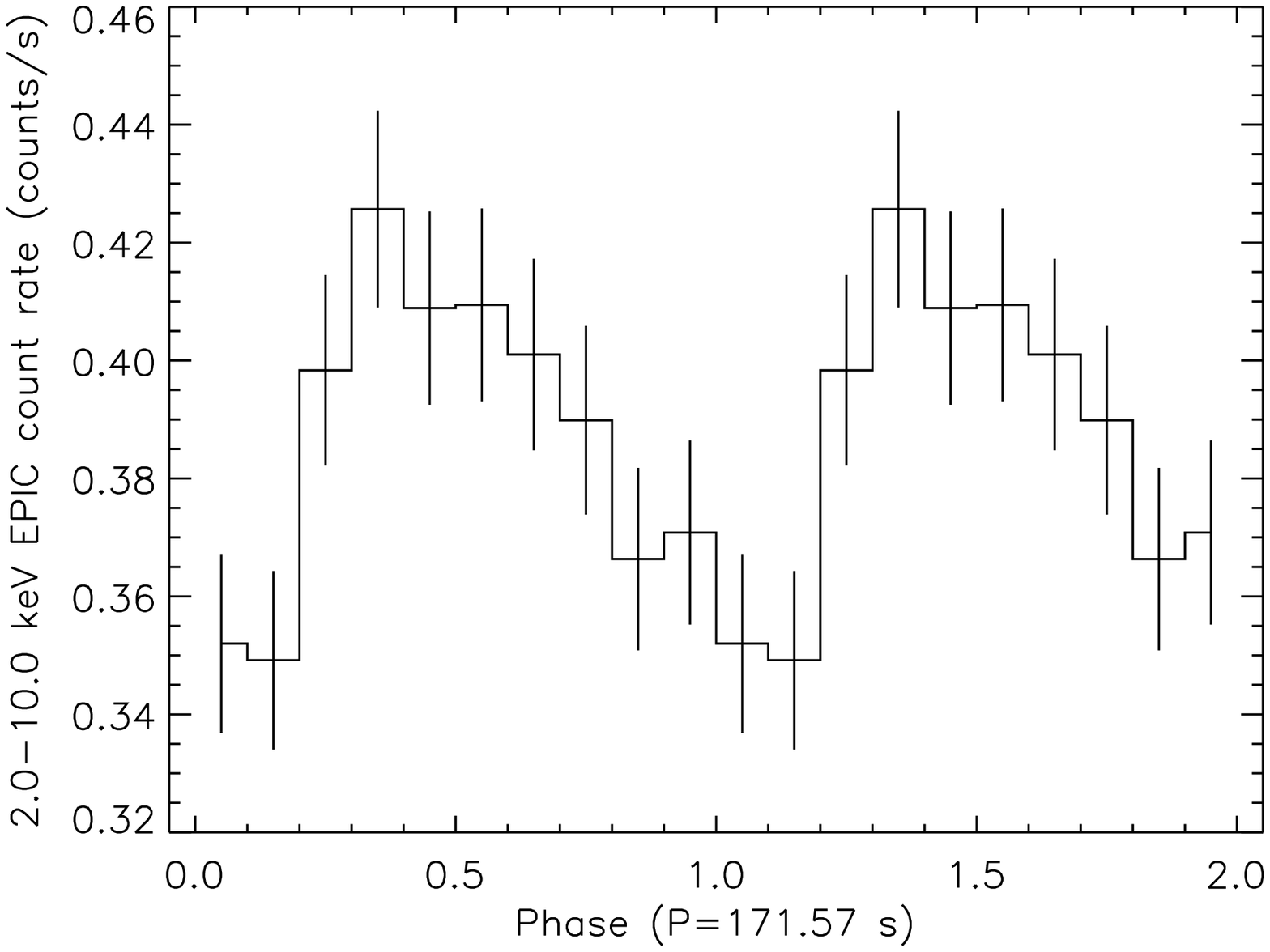}
\end{tabular}
\vspace*{-0.2cm}
\caption{Timing analysis of \ob\ in the 2--10\,keV energy band. 
{\it left}: Power spectrum of the EPIC 2.6\,s-binned light curve. 
The enlargement shows the X-ray pulsation candidate at the frequency 5.85$\pm$0.07\,mHz corresponding to a period of $171\pm2$\,s. {\it right}: Folded light curve  for the period $171.6\pm0.3$\,s found with periodogram analysis.} 
\label{fig:timing}
\vspace*{-0.2cm}
\end{figure*}
Therefore, the hydrogen column density estimated here for the absorbed
power law model is consistent with a location of this transient object  
at the distance of the Galactic center. 
We also test a thin thermal model {\tt mekal} but 
in case of solar metal abundance, we obtain an  
unphysically high temperature greater than 50\,keV. 
If the metal abundance is let as a free parameter, 
the temperature is lower with $kT=18.6^{+22.8}_{-7.0}$\,keV, 
and the upper limit of the metal abundance is about 0.3, 
due mainly to the lack of obvious spectral emission lines. 
Therefore the power-law model is our best fit.
We obtain an upper limit of the equivalent widths for a narrow iron K${\alpha}$
line ($\sigma$=10\,eV) between 6.4 (``neutral'') and 7\,keV (H-like) 
of 30--86\,eV. To look for a possible time variability of the spectrum, 
we split the observation into two parts, but we find no statistically 
significant variations of the photon index, the absorption column 
and the 2--10\,keV flux.

\section{Period analysis}\label{sec:timing}

 We have searched the combined EPIC data with a time bin of 2.6\,s for X-ray 
pulsations in the 2--10 keV range. The FFT power density spectrum 
 shows a peak at 5.85$\pm$0.07 mHz or 171$\pm$2\,s (Fig.~\ref{fig:timing} left).
The probability that the peak at this frequency is produced by just white 
photon noise is 2$\%$ has been found by Monte Carlo simulations. 
To cope with the possibility that the signal is not constant with just random 
fluctuations, we assessed the false alarm probability 
making use of the periodogram analysis  
(Lomb \cite{L76}, Scargle's \cite{S82}) in the form suggested by Andronov (\cite{A94}):  
$x(t)=a-R\cos(2\pi(t-T_0)/P)$, where $x(t)$ is the 2.6\,s-binned signal at time $t$, 
$a$ is phase-averaged mean value, 
$R$ is the semi-amplitude, $P$ is the trial period, and $T_0$ is an initial epoch.
 The MOS1, MOS2 and PN data as well as 
their sum have been analysed separately, and a candidate period 
at P = 171.57$\pm$0.25\,s has been found consistent with the period found by the FFT 
analysis. The candidate period has a signal to noise ratio (S/N) of 
2.9 (MOS1), 3.3 (MOS2), 6.0 (PN) and 9.7 (total). The S/N ratio is defined 
as the ratio of the count rate obtained for the candidate frequency divided 
by the count rate averaged over the full frequency band. Monte Carlo 
simulations were done for 50\,000 evenly spaced trial frequencies ranging from 
10$\sp{-5}$ s$\sp{-1}$ up to the Nyquist frequency of 1/2 s$\sp{-1}$, 
and resulted in a false alarm probability of 0.37 for 7638 independent frequencies.
We have also exercised a test on the phase relation among the three 
instruments by comparing the fitted values for T$\sb{0}$, which are
$T_{0,\rm MOS1}$=7678$\pm$11\,s, $T_{0,\rm MOS2}$=7682$\pm$11\,s, and $T_{0,\rm PN}$7665$\pm$8\,s.
The largest phase shift is 0.10$\pm$0.08 which suggests that there is no 
significant shift. In the context of this approach,    
we have estimated a false alarm probability, defined as the probability to get 
by chance the observed phase ``coincidence'', by computing 10$\sp{8}$ trial 
data sets with uniformly distributed phases. 
For each trial set, a weighted mean and variance is computed using
 the previous measured instrumental uncertainties.
The simulated variance of the phases exceeds  
the measured variance in 3.4$\%$ of all trials.   
In summary the false alarm probability for the suggested period
 is somewhere between 2 and 3.4$\%$ ignoring any modulation of the signal and 
could be around 37$\%$ otherwise.
Finally, we give the best fit parameters of the signal function for the data 
of  the three instruments combined: $a=0.368\pm0.005,$ $R=0.032\pm0.007,$ $T_0=7672.2\pm6.2$\,s 
(corresponding to mjd=2550.80667, i.e. October 3, 2002 at 7:21:36.0). 
The peak-to-peak amplitude of the periodic variations is $\sim16\%\pm2$,
which is the same for all three 
instruments within the error estimates.
Fig.~\ref{fig:timing} (right) shows the light curve folded with this period. 
We have extracted phase resolved spectra selecting data 
corresponding to peak (phase=0.2--0.6), and to trough (phase=0.6--1.2), and  
we found no significant spectral differences. 

\section{The nature of \ob}

We estimate the amplitude of the luminosity outburst from the deepest 
{\sl Chandra} observation of Sgr\,A*, 167\,ks of exposure obtained on
  May 25, 2002, i.e. four months before our {\sl XMM-Newton} observation.
 We found a 3-$\sigma$ count rate upper limit of 1.2 $\times$ 10$^{-4}$\,count\,s$^{-1}$ at the location of \ob~in the 0.5--10\,keV energy band.
 Assuming $\Gamma$=1.6 (see Table~\ref{tab:fit}),
 we find $L_{\rm X}$(0.5--10\,keV)$\le 1.2 \times 10^{32}\,(d/{\rm 8 kpc})^{2}$\,erg\,s$^{-1}$ during the quiescent state. 
Therefore, this source has exhibited a luminosity increase of a factor of 
about 1\,300
 over four months. In addition the source was not detected eight months
 after the present {\sl XMM-Newton} observation by {\it Chandra} during a
 short exposure (25\,ks) obtained on June 19, 2003
 leading to a 3-$\sigma$ 0.5--10\,keV count rate upper limit of 3.7 $\times$ 
10$^{-4}$\,count\,s$^{-1}$ at the location of \ob, 
corresponding to $L_{\rm X}$(0.5--10\,keV)$\le 3.7 \times 10^{32}\,(d/{\rm 8\,kpc})^{2}$\,erg\,s$^{-1}$.   
Then \ob~ is a transient source detected here, for the first time, during an outburst
 on October 3, 2002, ruling out for example the outburst of a supernova. 
 Such very rapid luminosity decrease is not compatible with a tidal
 disruption of a star as observed in some non active galaxies (e.g., Halpern et al. \cite{HGK04}).
 The X-ray properties suggests an X-ray binary nature, involving a
 compact object which could be either a white dwarf, or a neutron star, or a
 black hole.\\
\indent The typical outburst luminosity of white dwarf system 
is only up to about 10$^{34}$\,erg\,s$^{-1}$ (e.g., \object{GK Per}, Sen \& Osborne \cite{SO98}), 
implying a distance upper limit of about 2.5\,kpc, and the bulk of the observed 
hydrogen column density would be for instance intrinsic, produced by the accreting material. 
However, here the amplitude between the outburst phase and the quiescent phase is
 much larger than the one observed in cataclysmic variables. For example, GK Per, 
which has the most similar X-ray properties
 (e.g., $N_{\rm H}$, kT$_{\rm brems}$, period pulsation), has only an amplitude of 
about 10 (Hellier et al. \cite{H04}), and a significant Fe\,K${\alpha}$ line is seen 
during outburst with an equivalent width of 160$\pm$20\,eV typical for cataclysmic variables
 (see Ezuka \& Ishida \cite{EI99}).
 Therefore, a cataclysmic variable origin is very unlikely for \ob.\\ 
\indent The companion of the neutron star or of the black hole   
can be either a low mass object or a high-mass object.
A subclass of low mass X-ray binaries are transient systems, called {\sl X-ray novae}, 
which undergo sometimes outbursts. 
For most of the time, X-ray novae are in quiescent state, 
where the mass accretion rate from the disk to the compact object is very small,
producing a low-level of X-ray emission.
The quiescent 0.5--10\,keV X-ray luminosities of neutron star X-ray novae are about 
10$^{32}$--10$^{34}$\,erg\,s$^{-1}$, i.e., 100 times higher than those for black
hole X-ray novae ($\sim$10$^{30}$--10$^{33}$\,erg\,s$^{-1}$),
 as shown by Garcia et al.\ (\cite{G01}). 
Here we find an upper limit of the luminosity in the quiescent state of 
  1.2 $\times$ 10$^{32}\,(d/{\rm 8\,kpc})^{2}$\,erg\,s$^{-1}$, i.e., compatible with 
a black hole X-ray nova or a neutron star X-ray nova. 
The lower limit amplitude between the outburst phase and the quiescent phase is
consistent with both a neutron star and a black hole (Chen et al. \cite{C97}). 
Therefore \ob~ can be either a neutron star X-ray nova or a black hole X-ray nova.  
\ob~can also be a high mass X-ray binary formed by a neutron star or a black hole
 and a primary Be star companion.
Indeed, as reported by  Liu et al.\ (\cite{Liu00}), 
most of the transient high mass X-ray binaries are Be star systems. 
Roughly two-third of a sample of 130 high mass X-ray binaries are Be/X-ray binaries, 
and X-ray pulsations have been found in about 60 Be/X-ray binary systems
(Zi\'olkowski \cite{Z02}). 

We conclude that a higher $S/N$ X-ray observation of \ob~ during
 its outburst phase is needed 
to confirm or reject the candidate X-ray pulsation at 172\,s 
 suggested by our timing analysis. 
If confirmed, we will be able to firmly identify the compact object
 as a neutron star. Moreover near-infrared follow-up observations 
are needed to identify and to characterize the companion star.

\begin{acknowledgements}
This work is based on observations obtained with {\sl XMM-Newton}, 
an ESA science mission with instruments and contributions directly 
funded by ESA Member States and the USA (NASA).  
 D.P. is supported by a MPE fellowship.
\end{acknowledgements}


\end{document}